\documentclass[twocolumn]{emulateapj}
\newcommand{\fathi}{{\bf F10}}
\def\rd{$r_d$}

\def\kmsmpc{$\mbox{km s}^{-1}\mbox{ Mpc}^{-1}$}

\def\deg{^\circ}

\shorttitle{Revisiting the Scale Length--$\mu_0$ Plane and the Freeman Law}
\shortauthors{Kambiz Fathi}

\slugcomment{Accepted for publication in The Astrophysical Journal Letters}
\begin{document}

\title{Revisiting the Scale Length--$\mu_0$ Plane and the Freeman Law in the local universe}
\author{Kambiz Fathi \altaffilmark{1,2}}

\altaffiltext{1}{Stockholm Observatory, Department of Astronomy, Stockholm University, AlbaNova Center, 106 91 Stockholm, Sweden}
\altaffiltext{2}{Oskar Klein Centre for Cosmoparticle Physics, Stockholm University, 106 91 Stockholm, Sweden}

\begin{abstract}
We have used Virtual Observatory technology to analyse the disk scale length \rd\ and central surface brightness $\mu_0$ for a sample of 29955 bright disk galaxies from the Sloan Digital Sky Survey. We use the results in the $r$-band and revisit the relation between these parameters and the galaxy morphology, and find the average value $\langle\mu_0\rangle=20.2\pm0.7$ mag~arcsec$^{-2}$. We  confirm that late type spirals populate the lower left corner of the \rd--$\mu_0$ plane and that the early and intermediate spirals are mixed in this diagram, with disky ellipticals at the top left corner. We further investigate the Freeman Law and confirm that it indeed defines an upper limit for $\mu_0$ in bright disk galaxies with $r_\mathrm{mag}<17.0$, and that disks in late type spirals ($T \ge 6$) have fainter central surface brightness. Our results are based on a volume corrected sample of galaxies in the local universe ($z < 0.3$) that is two orders of magnitudes larger than any sample previously studied, and deliver statistically significant implications that provide a comprehensive test bed for future theoretical studies and numerical simulations of galaxy formation and evolution.
\end{abstract}


\section{Introduction}
\label{sec:intro}
The current mainstream galaxy formation paradigm states that galaxy disks form within dark matter halos and that there is an intimate relation between the scale length \rd\ of the disk and that of the halo. The mass distribution of the disk is entirely set by the \rd\ and for example in the exponential case, 60\% of its total mass is confined within two scale lengths and 90\% within four scale lengths. Moreover, the angular momentum of the disk is set by \rd\ and the mass distribution of its host halo, and the fact that the angular momentum vectors are aligned suggests that there is a physical relation between the two. During the formation process, galaxy mergers and associated star formation and feedback processes play a crucial role in the resulting structure, however, the observed sizes of disks suggest that the combination of these physical processes yield that galactic disks have not lost much of the original angular momentum acquired from cosmological torques \citep[e.g., ][]{WR78, FE80}. The hierarchical and infall models predict comparable \rd\ and in a cold collapse scenario \citep{Vand59} since angular momentum is conserved, immediately after the collapse the gas is supported by rotation so that it quickly collects in the region where a disk forms with substantially higher rotation velocity than the halo is to form. A large \rd\ disk forms when the disk mass is smaller than the halo mass over the disk region, and vice versa, a small \rd\ disk forms when the mass of the disk dominates the mass of the halo in any part of the disk. The self gravitating disk will also modify the shape of the rotation curve near the centre of a galaxy \citep{GS99} and the disk is then set to undergo secular evolution. The natural implication of this scenario is that the \rd\ dictates the life of a disk, and consequently, is a prime factor which determines the position of a galaxy on the Hubble sequence. 

One prominent indicator for a smooth transition from spiral toward S0 and disky ellipticals is provided by the \rd--$\mu_0$ diagram where $\mu_0$ is the central surface brightness of the disk, where spirals and S0s are mixed and disky ellipticals populate the upper left corner of this diagram \citep{Kent85,SB95}. Another instructive relation is the Freeman Law \citep{Freeman70} which relates $\mu_0$ to the galaxy morphological type. Although, some studies have found that the Freeman Law is an artefact due to selection effects \citep[e.g., ][]{Disney76, Bothun81, SvdB98}, recent work have shown that proper consideration of selection effects can be combined with kinematic studies to explore an evolutionary \citep[e.g., ][]{vdKruit87, deJong96, KSW00, vdKruit02}.

In the comparison between theory and observations, two issues complicate matters. On the theory side mapping between initial halo angular momentum and \rd\ is not trivial, partly due to the fact that commonly the initial specific angular momentum distribution of the visible and dark component favour disks which are more centrally concentrated disks than exponential \citep[e.g., ][]{deJL00, vdeBosch01}. Observationally, comprehensive samples have yet not been studied, and the mixture of different species such as low and high surface brightness galaxies complicate the measurements of disk parameters \citep{McGaugh95, GdeB01}. 

Here we analyse the \rd\ and $\mu_0$ from an unprecedentedly large sample of  bright disk galaxies in the nearby universe ($z< 0.3$) using the Sloan Digital Sky Survey (SDSS) Data Release 6 \citep{Yorketal00, AMcetal08}. As detailed in \citet{Fathi10} (hereafter \fathi), both parameters were robustly determined for 30374 galaxies in the $r$-band  (only 29955 used here as described in section~\ref{sec:data}), whereas in other SDSS bands, the derived parameters are valid only for subsets of this sample. In the $g$, $i$, and $z$-band ($\approx 27000$--30000 galaxies) the sample sizes are comparable to the one presented here. In the $u$-band, \rd\ and $\mu_0$ were robustly derived for a few hundred objects and therefore, not considered here. Throughout this paper, unless otherwise stated, we use disk parameters in the $r$-band and investigate the two relations mentioned above, in order to provide a comprehensive test bed for forthcoming cosmological simulations (or analytic/semi-analytic models) of galaxy formation and evolution.

\section{The Derived Parameters and Sample Correction}
\label{sec:data}
\begin{figure}
\begin{center}
\includegraphics[width=.49\textwidth]{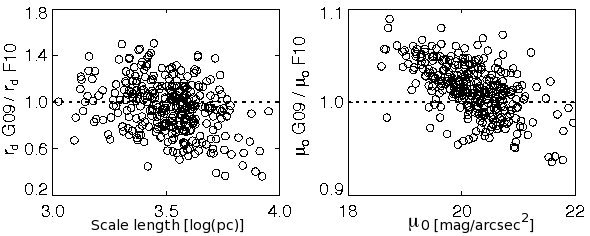}
  \caption{Comparison of our $r$-band \rd\ and $\mu_0$ with \citet{G09}, where the ratios between the two studies are shown in the corresponding panel ($y$-axis).} 
\label{fig:fathigadotti}
\end{center}
\end{figure}

The derivation of \rd\ and $\mu_0$ for the SDSS sample is detailed in \fathi, and testing the code with standard IRAF routines delivered fully consistent results. Comparing disk parameters with those from several published samples \citep{deJong96, G03, A04, K06, Betal07} failed to deliver any objects in common. However, cross matching with the multi-band image fitting and multi-component (disk, bar, bulge, and central source) profile decomposition of nearly 1000 galaxies based on SDSS \citep{G09} returned 322 disk galaxies for which we derived comparable values (see Fig.~\ref{fig:fathigadotti}). As outlined in \fathi, we made use of the SkyView services to obtain cut-outs from the SDSS tiles, with the galaxy counts derived in the reference frame of the tile that is at the centre of the SkyView cut-out. The disk central surface brightness can thus be derived using the procedure provided by the SDSS services.
Applying pixel area correction and Tolman cosmological dimming correction $(z+1)^4$ we were able to obtain the $\mu_0$ in correct units of mag arcsec$^{-2}$. The longitudes provided by the SDSS services show that for our sample $\|b\|>20\deg$, and we use the standard extinction correction law of \citet{Getal94} to correct for this effect. Following the description in \citet{V90} and \citet{deJong96}, we use a range of internal extinction parameters (0 for optically thick galaxies to 1 for transparent galaxies) for every diagram to quantify this effect throughout our analysis, but present diagrams calculated with the extinction parameter of 0.26 \citep{deJong96}.

We also derived the asymmetry parameter and the calculation of the inverse concentration parameter (\fathi). By examining a small subset of galaxies for which reliable morphological classifications are available form the LEDA catalogue \citep{Petal03}, we assert that the stellar velocity dispersion, asymmetry, and concentration can be used with caution as indicators of the position of a galaxy along the Hubble sequence. We stress, however, the data points have large spread and at best have a correlation coefficient of determination $\mathfrak{R^2}=0.31$, well below one-sigma level (see \fathi\ for details).

Our derived \rd\ and $\mu_0$ for the SDSS sample will have to be corrected for the frequency of galaxies with a certain property in a volume to be included in the sample. The volume over which we can detect low luminosity galaxies is smaller than the volume over which we can detect high luminosity galaxies. In order to calculate the number density contribution of each galaxy, we calculate the maximum volume over which it is observable $V_\mathrm{max}$. We follow the volume weighting recipe described in \citep[e.g., ][]{vdKruit87, deJong96}, and weight each galaxy by the calculated $V_\mathrm{tot}/V_\mathrm{max}$, where $V_\mathrm{tot}$ is the volume contained within the distance to farthest galaxy in the sample \citep[e.g., ][]{Davies90, B03}. To calculate the volumes, we assume a $\Lambda$CDM Universe with the Hubble parameter $H_0=70$ \kmsmpc, $\Omega_\Lambda = 0.7$, and $\Omega_M = 0.3$, and for this calculation use $g$-band luminosities to calculate luminosity distances as these are more representative of $B$-band values. Changing the cosmology has no significant effect on the analysis carried out here. 

\begin{figure}
\begin{center}
 \includegraphics[width=.49\textwidth]{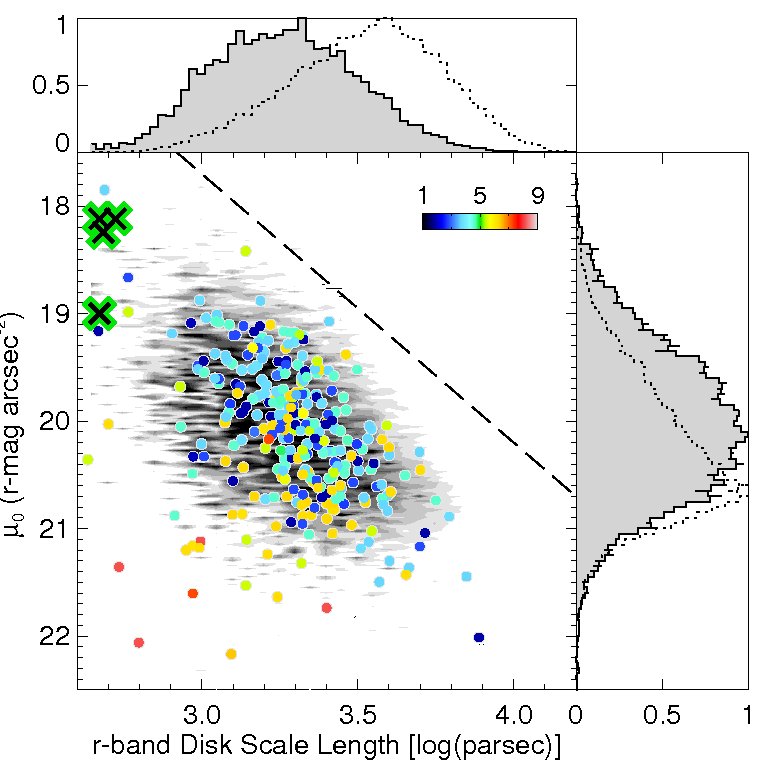}
  \caption{The \rd--$\mu_0$ plane of our sample (density plot of the 29955 galaxies), with the 282 reliably classified galaxies illustrated as coloured points where the colour indicates the numeric morphological type T. The dashed line (slope 2.5) illustrates the line of constant disk luminosity, and the crosses at the top right illustrate the seven disky ellipticals from \citet{SB95}. At the top and right, the normalised distribution of each parameter is shown before (dotted) and after (solid) applying the volume correction described in section~\ref{sec:data}.} 
\label{fig:mu0rd}
\end{center}
\end{figure}

This method has the disadvantage that if the near part of the sample is overdense, where a magnitude-limited survey is most sensitive to low-luminosity galaxies, then the $V_\mathrm{tot}/V_\mathrm{max}$ will become very large for the faintest objects. To avoid this problem, we make a cut in the sample by removing galaxies with unreasonably large weights ($>10000$). This leaves us with 29955 galaxies at $z<0.3$ and $r_\mathrm{mag}<17.0$, for which we obtain volume corrected parameter distributions (see e.g., Fig.~\ref{fig:mu0rd}).

\begin{figure*}
\begin{center}
 \includegraphics[width=.99\textwidth]{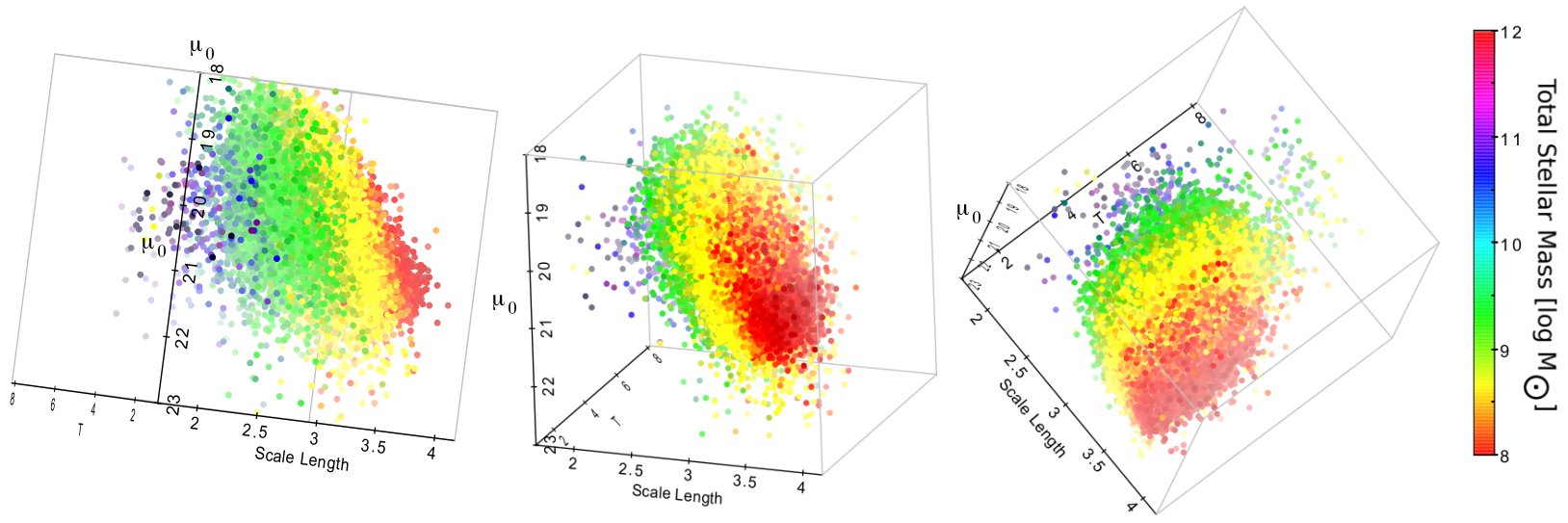}
  \caption{Four-dimensional parameter space of $r_d$, $\mu_0$, morphological type T, and total stellar mass. Three different views are illustrated here, the $\mu_0$ values are given in mag~arcsec$^{-2}$ and \rd\ and mass are given in decimal logarithm of parsecs and solar masses, respectively.} 
\label{fig:mu0rdTypeMass}
\end{center}
\end{figure*}
\begin{figure}
\begin{center}
 \includegraphics[width=.49\textwidth, height=12cm]{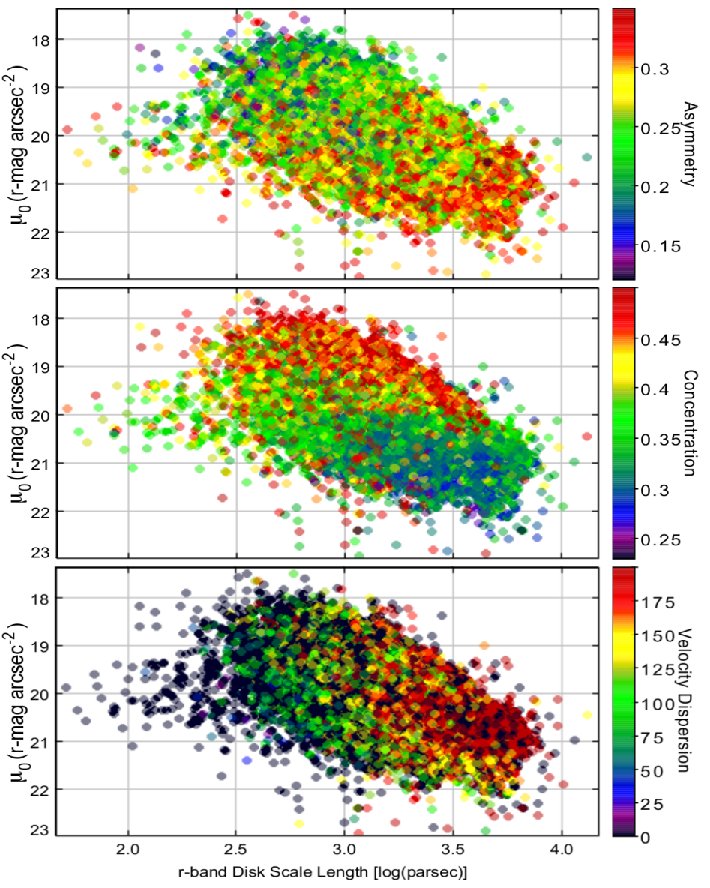}
  \caption{The \rd--$\mu_0$ plane colour coded by asymmetry (top), concentration parameter (middle), and velocity dispersion $\sigma$ (bottom), where the ranges have been chosen to amplify the effects discussed here.} 
\label{fig:mu0rdCASigma}
\end{center}
\end{figure}
\section{The \rd--$\mu_0$ Plane}
\label{sec:rdmu0}
Figure~\ref{fig:mu0rd} shows that the derived \rd\ and $\mu_0$ follow the linear trend previously discussed in the context of the formation and evolution of disks \citep{deJong96,Courteau96,vdeBosch98,GdeB01}. We find that $\mu_0$ has a Gaussian distribution with $\langle\mu_0\rangle=20.1\pm0.7$ mag~arcsec$^{-2}$ with a linear trend seen in Fig.~\ref{fig:mu0rd} (applying different internal extinction parameters changes this mean value by 0.2 mag~arcsec$^{-2}$). The top right corner is enclosed by the constant disk luminosity line, void of objects. The top right corner is also the region where the disk luminosity exceeds $3L^\star$, thus the absence of galaxies in this region cannot be a selection effect since big bright galaxies cannot be missed in our diameter selected sample. However, it is clear that selection effect plays a role in populating the lower left corner of this diagram. Analogue to the \rd--$\mu_0$ plane, the Tully-Fisher relation implies lines of constant maximum speed a disk can reach. 

The type dependency of the \rd--$\mu_0$ plane was previously presented by \citet{GdeB01} who used 120 high and low surface brightness galaxies to explore the type-dependency of the diagram. Our 282 well-classified galaxies (illustrated with coloured dots in Fig.~\ref{fig:mu0rd}) follow the results of \citet{GdeB01} and \citet{G09}. Disks of intermediate and early type spirals have higher $\mu_0$ while the late type spirals have lower $\mu_0$, and they populate the lower left corner of the diagram. The \rd, on the other hand, does not vary as a function of morphological type (\fathi). 

Figure~\ref{fig:mu0rdTypeMass} shows that a forth quantity is equally important in this analysis. The total galaxy mass obtained from \citet{Kauffmannetal03} and \citet{Brinchmannetal04}, separates the data along lines parallel to the dashed lines drawn in Fig.~\ref{fig:mu0rd}. This is indeed also confirmed by the Tully-Fisher relation. Moreover, we validate that the lower mass galaxies are those with type $\ge 6$. 

Our data also show the expected trends: Asymmetry and concentration increase for later types, and central stellar velocity dispersion decrease for later type spiral galaxies, however, we note that these correlations are well below one-sigma confidence level (see end of section~\ref{sec:data} and \fathi). In Fig.~\ref{fig:mu0rdCASigma} we illustrate \rd--$\mu_0$ plane where the data points have been colour coded by asymmetry, concentration, and velocity dispersion as a rough indication of the galaxies in different types. 

The top panel of Fig.~\ref{fig:mu0rdCASigma} indicates that the higher asymmetry galaxies populate a region more extended toward the bottom right corner with respect to the high asymmetry galaxies. The middle panel shows an opposite trend, and the bottom panel shows that larger velocity dispersion has the same effect as asymmetry.  In full agreement with \citet{GdeB01}, Fig.~\ref{fig:mu0rdCASigma} confirms that galaxies with early and intermediate morphologies are mixed along a linear slope of 2.5 in the \rd--$\mu_0$ plane, coinciding with the region populated by S0s as shown by \citet{Kent85} and disky ellipticals shown by \citet{SB95}.

\begin{figure}
\begin{center}
\includegraphics[width=.49\textwidth]{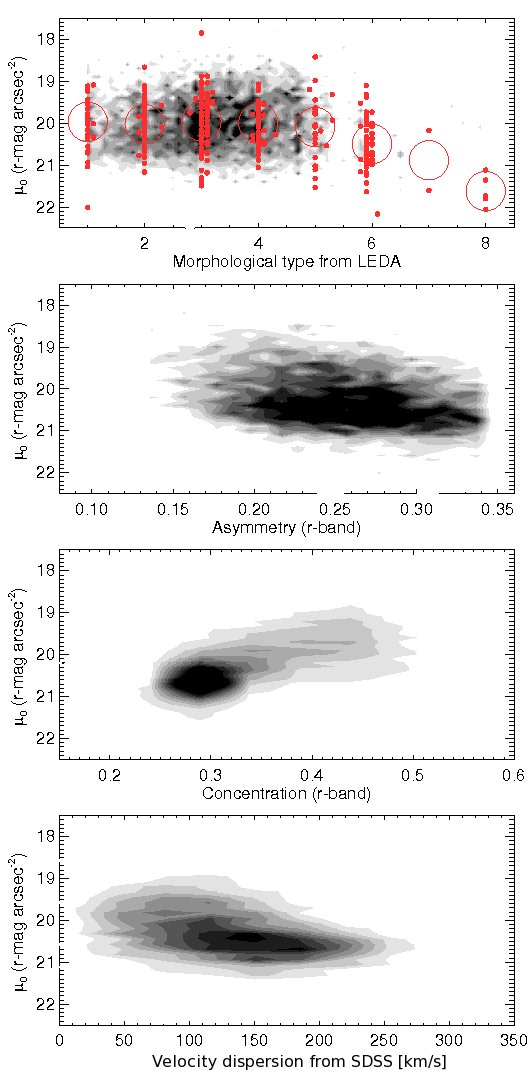}
  \caption{Volume corrected distribution of disk central surface brightness $\mu_0$ versus morphological type T, asymmetry, concentration parameters, and stellar velocity dispersion. The red dots illustrate the 282 galaxies for which robust morphological types were given in the LEDA catalogue (type error smaller than 0.5) and the open circles show the average $\mu_0$ for each type.} 
\label{fig:mu0type}
\end{center}
\end{figure}
\section{The Freeman Law}
\label{sec:freeman}
We plot the Freeman (1970) diagram, where the density plot indicates that disks have constant surface brightness put to Scd galaxies ( types $T = 5$, top panel of Fig.~\ref{fig:mu0type}). The volume corrected density plot also shows that the vast majority of the sample is of morphological type earlier than 5, thus introducing a bias in the observed trend. When we consider only the galaxies with robust morphological classification (red dots in the top panel of Fig.~\ref{fig:mu0type}), we note that $\mu_0$ drops for galaxies of $T\ge6$ \citep[c.f., Fig.~3 of ][]{deJong96} with the significant advantage that our sample contains five times more galaxies than the number of late type spirals of \citet{deJong96}. Furthermore, using the entire sample of 29955 galaxies, we confirm that the Freeman Law indeed defines an upper limit for $\mu_0$ in bright disk galaxies at $z<0.3$ (low-surface brightness galaxies are not present in our sample). Combined with our previous results, i.e. that \rd\ varies by two orders of magnitude independent of morphological type, this result implies that disks with large scale lengths (and typically higher mass) not necessarily have higher $\mu_0$. 

We use the asymmetry, concentration, and stellar velocity dispersion as indicators of morphological type. The lower three panels of Fig.~\ref{fig:mu0type} illustrate that the density plots for the entire sample and the galaxies. None of the trends are statistically significant in exploring these parameters against the Freeman diagram. The lack of strong correlation may well be attributed to the fact that three parameters that we have used here do not correlate strongly with morphological type. The concentration parameter is further complicated by the nature of the bulges \citet{G09}. We conclude that these parameters cannot be used to further explore the Freeman Law.

\section{Conclusions}
\label{sec:conclusions}
We have derived and analysed \rd\ and $\mu_0$ in the $r$-band from SDSS images of the largest ever volume corrected sample of disk galaxies in the local universe ($z< 0.3$ and $r_\mathrm{mag}<17.0$). We confirm that early and intermediate type spirals are mixed in the \rd--$\mu_0$ plane and late type spirals ($T\ge 6$) populate the lower left corner of this plane. Varying the internal extinction value does not change the trends found here, and the average value $\langle\mu_0\rangle=20.1\pm0.7$ mag~arcsec$^{-2}$ only varies slightly. Moreover, the Freeman Law defines an upper limit for $\mu_0$, where disk galaxies of type $T\ge6$ have fainter $\mu_0$. 

These results in $r$-band are comparable with those in other SDSS bands, and the average $\mu_0$ in other bands are: $g$-band ($20.3\pm0.7$ mag~arcsec$^{-2}$), $i$-band ($20.1\pm0.7$ mag~arcsec$^{-2}$ and fully consistent with the median value from \citet{G09}), and $z$-band ($21.6\pm0.7$ mag~arcsec$^{-2}$). Given the intrinsic scatter in each band, we do not find these differences significant to interpret the effects of dust and/or stellar populations \citep[c.f., ][]{Petal94} moreover that presented in \fathi. Furthermore, our results imply that disks with large scale lengths (and typically higher mass) not necessarily have brighter $\mu_0$. 

As earlier discussed by e.g., \citet{Kent85,vdKruit87,SB95,vdKruit02} and many others, selection effects could cause artificially tight or incorrect correlations. In the two relations analysed here, we find typically larger scatter than previous analyses, and although our sample represents bright disks, the sample size adds credibility to our findings. These results are fully consistent with the common understanding of the \rd--$\mu_0$ plane and the Freeman Law, and they contribute to past results since they are based on a sample which is two orders of magnitude greater than any previous study, with more than five times more late type spiral galaxies than any previous analysis.

\section*{Acknowledgments}
This work made use of EURO-VO tools and services, funded by the European Commission through contract number 212104 (AIDA  under FP~7). Also SDSS services, NASA's SkyView, HyperLeda, and the TOPCAT software (http://www.star.bristol.ac.uk/$\sim$mbt/topcat) were used. I acknowledge support from the Swedish Research Council (Vetenskapsr\aa det) and the hospitality of ESO-Garching where parts of this work were carried out. I am indebted to Evanthia Hatziminaoglou for her help with the use of VO tools, thank Peter Erwin and David Wilman for their valuable input on the cosmological volume calculation, and thank Dimitri Gadotti for kindly sharing his derived scale lengths. Finally, I am grateful to the anonymous referee for invaluable comments which improved the quality of this work.

\label{lastpage}
\end{document}